\documentclass[a4paper, 11pt, oneside]{article}

\usepackage{amsmath}
\usepackage{amsfonts}
\usepackage{comment}
\usepackage{epsfig}
\usepackage[font=small,labelfont=bf]{caption}
\usepackage[table]{xcolor}
\usepackage{tabularx}
\usepackage{amsthm}
\usepackage{url}
\usepackage{fullpage}
\usepackage{booktabs}
\usepackage{caption}

\usepackage{setspace}
\usepackage{multirow}
\usepackage{multicol}

\usepackage{enumerate}
\usepackage{float}
\usepackage[stable]{footmisc}
\usepackage{titletoc}
\usepackage{caption}

\usepackage{setspace}

\usepackage[left=1in, right=1in, top=1in, bottom=1in, includefoot, headheight=13.6pt]{geometry}

\usepackage{titlesec}

\def\P{\hbox{\it I\hskip -2pt P}}
\def\E{\hbox{\it I\hskip -2pt E}}
\def\N{\hbox{\it I\hskip -2pt N}}

\usepackage{graphicx}

\usepackage[nottoc]{tocbibind}

\usepackage{textcomp}
 
\title{The impact of systemic risk\\[-1ex] on the diversification benefits of a risk portfolio}

\date{}

\author{\Large Marc Busse $^{1}$, Michel Dacorogna $^{1}$, Marie Kratz $^{2},$*\\[1ex]
\small $^{1}$ SCOR SE, Switzerland\\
\small $^{2}$ ESSEC Business School Paris, CREAR Risk Research Center
}

\begin{document}
\setlength{\parindent}{0in}
\parskip 2.0ex  % skip some length before paragraph
\onehalfspacing
\maketitle
\begin{abstract}
\noindent
Risk diversification is the basis of insurance and investment. It is thus crucial to study the effects that could limit it. One of them is the existence of systemic risk that affects all the policies at the same time. We introduce here a probabilistic approach to examine the consequences of its presence on the risk loading of the premium of a portfolio of insurance policies. This approach could be easily generalized for investment risk. We see that, even with a small probability of occurrence, systemic risk can reduce dramatically the diversification benefits. It is clearly revealed via a non-diversifiable term that appears in the analytical expression of the variance of our models. We propose two ways of introducing it and discuss their advantages and limitations.
By using both VaR and TVaR to compute the loading, we see that only the latter captures the full effect of systemic risk when its probability to occur is low.\\

\vspace{\fill}
{\emph{\hspace*{-18pt} 2000 AMS classification}: 91B30; 91B70; 62P05\\[1ex]
* Marie Kratz is also member of MAP5, UMR 8145, Univ. Paris Descartes }\\
\end{abstract}

\section{Introduction}\label{sec-intro}

Every financial crisis reveals the importance of systemic risk and, as a consequence, the lack of diversification. Diversification is a way to reduce the risk by detaining many different risks, with various probabilities of occurrence and  a low probability of happening simultaneously. Unfortunately, in times of crisis, most of the financial assets move together and become very correlated. The 2008/09 crisis is not an exception. It has highlighted the interconnectedness of financial markets when they all came to a stand still for more than a month waiting for the authorities to restore confidence in the system (see e.g. the {\it Systemic Risk Survey of the bank of England} available on line). For any financial institution, it is important to be aware of the limits to diversification, while, for researchers in this field, studying the mechanisms that hamper diversification is crucial for the  understanding of the dynamics of the system (see e.g. \cite{caba}, \cite{contAl} and references therein). Both risk management and research on risk would enhance our capacity to survive the inevitable failures of diversification. A small fact, like turmoils in the sub-prime in the US real-estate market, a relatively small market compared to the whole US real-estate market, can trigger a major financial crisis that extends to all markets and all regions in the globe. Systemic risk manifests itself by a breakdown of the diversification benefits and the appearance of dependence structures that were not deemed important during normal times.

In this study, we introduce a simple stochastic modelling to understand and point out the limitations to diversification and the mechanism leading to the occurrence of systemic risk.
The idea is to combine two generating stochastic processes that, through their mixture, produces in the resulting process a non-diversifiable component, which we identify to a systemic risk.
Depending on the way of mixing these processes, the diversification benefit appears with various strengths due to the emergence of the systemic component.
The use of such a model, which is completely specified, allows us to obtain analytical expressions for the variance, and then to identify the non-diversifiable term. With the help of Monte Carlo simulations, we explore the various components of the model and check that we reproduce the analytical results.

 The paper is organized as follows: in a first section, we introduce, to measure the effects on diversification, the standard insurance framework for pricing risk and define the risk measures that are used in this study. The second section is dedicated to the mathematical presentation of the model and its various approaches to systemic risk, as well as numerical applications. The obtained results are compared numerically and analytically in the third section, where we also discuss the influence of the choice of risk measure on the diversification benefits. We conclude the study and suggest new perspectives to extend it.

\section{Insurance framework}\label{sec-frame}

Before moving to stochastic modelling, let us introduce the insurance framework in which we are going to compute the risk diversification. It is an example of application, but  this study on systemic risk  could easily be generalized to any financial institution.

\subsection{The technical risk premium}\label{ss-techPremium}

In insurance, risk is priced based on the knowledge of the loss probability distribution.  Let $L$ be the random variable (rv) representing a loss defined on a probability space $(\Omega,{\cal A},\P)$.

\begin{itemize}
\item {\it One policy case}\\
For any policy incurring a loss $L^{(1)}$, we can define, as in \cite{Dac:Hu}, the technical premium, $P$, that needs to be paid,  as:
\begin{equation} \label{eq:premium}
P = \E[L^{(1)}] + \eta K + e \quad \text{with}
\end{equation}
$\eta$: the return expected by shareholders before tax\\
$K$: the risk capital assigned to this policy \\
$e$:  the expenses incurred by the insurer to handle this case.

An insurance is a company in which we can invest. Therefore the shareholders that have invested a certain amount of capital in the company expect a return on investment. So the insurance firm has to make sure that the investors receive their dividends, which corresponds to the cost of capital the insurance company must charge on its premium.
This is what we have called $\eta$.

We will assume that the expenses are a small portion of the expected loss
$$
e=a\E[L^{(1)}] \quad \text{with} ~0<a<\!\!<1\; ,
$$
 which transforms the premium as
\begin{equation} \label{eq:premium2}
P = (1+a) \E[L^{(1)}] + \eta K
\end{equation}
We can now generalize this premium principle  \eqref{eq:premium2}  for $N$ similar policies (or contracts).
\item {\it Case of a portfolio of $N$ policies}\\
The premium for one policy in the portfolio, incurring now a total loss $L^{(N)}=N\times L^{(1)}$, can then be written as
$$
P =\frac{ (1+a) \E[L^{(N)}] + \eta K_N}{N}=(1+a) \;\E[L^{(1)}] + \eta\;\frac{K_N}{N}
$$
where $K_N$ is the capital assigned to the entire portfolio.

\end{itemize}

\subsection{Cost of Capital and Risk Loading} \label{ssec-CoC}

First we have to point out that the role of capital for an insurance company is to ensure that the company can pay its liability  even in the worst cases, up to some threshold.
For this, we need to define the capital we have to put behind the risk. We are going to use a risk measure, say $\rho$,  defined on the loss distribution.
It  allows to estimate the capital needed to ensure payment of the claim up to a certain confidence level. We then define the risk-adjusted-capital $K$ as a function of the risk measure $\rho$ associated to the risk\footnote{We use here the word "risk" instead of "loss". In fact, these two words are used for one another in an insurance context.} $L$ as
\begin{equation} \label{eq:RAC}
K=\rho(L)-\E[L]
\end{equation}
since the risk is defined as the deviation from the expectation.

Note that we could have also defined $K$ as $K=\rho(L)-\E[L]-P$ since the premiums can serve to pay the losses. It would change the premium $P$ defined in \eqref{eq:premium} into $\hat{P}$ defined by
\begin{equation*} %\label{eq:altprem}
\hat{P}=\frac{1+a-\eta}{1+\eta}~\E[L]+\frac{\eta}{1+\eta} \rho(L)
\end{equation*}
Such an alternative definition would reduce the capital but does not change fundamentally the results of the study.

Consider a portfolio of $N$ similar policies, using the notation for the loss as in \S \;\ref{ss-techPremium}. Let $R$ denote the risk loading per policy, defined as  the cost of the risk-adjusted-capital per policy. Using \eqref{eq:RAC}, $R$ can be expressed as a function of the risk measure $\rho$, namely
\begin{equation}\label{eq: riskload}
R = \eta~\frac{K_N}{N}= \eta~ \biggl(\frac{\rho(L^{(N)})}{N} - \E[L^{(1)}] \biggr)
\end{equation}

\subsection{Risk Measures} \label{ssec-riskmeasure}

We will consider for $\rho$ two standard risk measures, the Value-at-Risk (VaR)  and the  tail Value-at-Risk (TVaR). Let us remind the definitions of these quantities (see e.g. \cite{McN:FE}).

The Value-at-Risk with a confidence level $\alpha$ is defined for a risk $L$ by
\begin{equation}
{\rm VaR}_{\alpha}(L) =inf\{q \in \mathbb{R} : \P(L>q)\le 1-\alpha\}  =inf\{q\in \mathbb{R} : F_{L}(q)\ge \alpha\}
\label{eq:VaR}
\end{equation}
where  $q$ is the level of loss that corresponds to a $\rm VaR_{\alpha}$ (simply the quantile of $L$ of order $\alpha$), and $F_L$ the cdf of $L$.

The tail Value-at-Risk at a confidence level $\alpha$ of $L$ satisfies
\begin{equation*}
{\rm TVaR}_{\alpha}(L)=\frac{1}{1-\alpha}\int_\alpha^1 {\rm VaR}_{u}(L)\mathrm{d}u \underset{F_L\;\text{contin.}}{=} \E[L\;|\; L>VaR_{\alpha}(L) ]
% = \frac{1}{1-\alpha}\int_\alpha^1 q_{u}(L)\mathrm{d}u
\end{equation*}
% where $q_u(L)={\rm VaR}_u(L)$.
When considering a discrete rv $L$, it can be approximated by a sum,
which may be seen as the average over all losses larger than $\rm VaR_\alpha$:
\begin{equation}
{\rm TVaR}_{\alpha}(L)=\frac{1}{1-\alpha}\sum_{u_i \geq \alpha}^1 q_{u_i}(L)\Delta u_i
\label{eq:TVaR}
\end{equation}
where $q_{u_i}(L)={\rm VaR}_{u_i}(L)$ and $\Delta u_i \equiv u_i - u_{i-1}$ corresponds to the probability mass of the particular quantile $q_{u_i}$.

\section{Stochastic modelling}\label{sec-model}

Suppose an insurance company has underwritten $N$ policies of a given risk.  To price these policies, the company must know the underlying probability distribution of this risk, as seen in the previous section.
In this study, we assume that each policy is exposed $n$ times to this risk.

In a portfolio of $N$ policies, the risk may occur $n\times N$ times. So, we introduce a sequence $(X_i, i=1,\ldots,Nn)$ of rv's $X_i$ defined on  $(\Omega,{\cal A},\P)$ to model the occurrence of the risk, with a given severity $l$. Note that we choose a deterministic severity, but it could be extended to a random one, with a specific distribution.

Hence the total loss amount, denoted by $L$, associated to this portfolio is given by
$$
L=l\,S_{Nn} \quad \text{with} \quad S_{Nn}:= \sum_{i=1}^{Nn} X_i
$$

\subsection{A first model, under the iid assumption}\label{iid}

We start with a simple model, considering a sequence of independent and  identically distributed (iid) rvs $X_i$'s. Let  $X$ denote the parent rv and
assume that it is simply Bernoulli distributed, {\it i.e.} the loss $lX$ occurs with some probability $p$:
$$
X= \left\{
\begin{array}{ll}
1 & \text{with probability}\quad p\\
0 & \text{with probability}\quad 1-p\
\end{array}
\right.
$$
Recall that $\E[X]=p$ and $var (X)=p(1-p)$.

Hence the total loss amount $L=l\,S_{Nn}$ of the portfolio is modeled by a binomial distribution ${\cal B}(Nn,p)$:
\begin{equation} \label{eq:cdf-L}
\P[L=k]=l \binom{Nn}{k} p^k (1-p)^{Nn-k}, \quad \text{for} ~ k=0,\cdots,Nn
\end{equation}
with  $\displaystyle \E[L]=\sum_{i=1}^{Nn} \E[X_i]= lNnp$, and,  by independence, $var(L)=lNn\; var(X)=lnp(1-p)$.

We are interested in knowing the risk premium the insurance will ask to a customer if he buys this insurance policy. So we compute the cost of capital given in~\eqref{eq: riskload} for an increasing number $N$ of policies in the portfolio, which becomes for this model:
\begin{equation}\label{eq: riskloadBernoulli}
R = \eta~ \biggl(\frac{\rho(L)}{N} - lnp \biggr)
\end{equation}
since the notation $L^{(N)}$ in \eqref{eq: riskload}  has been simplified to $L$ in this section.

Note that the relative risk per policy defined by the ratio $R/\E[L^{(1)}]$ is given by
\begin{equation}\label{eq: relriskBernoulli}
\frac{R}{\E[L^{(1)}]}= \eta~ \biggl(\frac{\rho(L)}{lnp} - 1 \biggr)
\end{equation}

%%%
{\it Numerical application.}

We compute the quantities of interest by fixing the various parameters.
We choose for instance the number of times one policy is exposed to the risk to be $n=6$.
Then we take as cost of capital $\eta=15$\%, which is a reasonable value given the fact that  the shareholders will obtain a return on investment after taxes of approximately 10\%, when considering a standard tax rate of 30\%.  For a discussion on the choice of the value of $\eta$, we refer to \cite{Dac:Hu}.
The unit loss $l$ will be fixed to $l=10$.
We choose in the rest of the study  $\alpha=99$\% for the threshold of the risk measure $\rho$.

We present in Table\;\ref{tbl-fairdist} an example of the distribution of the loss $L=lS_{1n}$ for one policy ($N=1$) when taking e.g. $p=1/n=1/6$ (same probability for each of the $n$ exposures). This would be the typical distribution for the outcome of a particular value when throwing a dice (see \cite{3M}).\\

\begin{table}[h]
\begin{center}
\caption[Loss Distributions]{{\small \sf The loss distribution for one policy with $n=6$ and $p=1/6$.}\label{tbl-fairdist}}
\begin{tabular}{c c@{\extracolsep{3em}} c@{\extracolsep{3em}} c}\hline
  &&&\\
 number  of losses & Policy Loss  &  Probability Mass   &     Cdf\\
       $k$     & $l~X(\omega)$ &     $\P[S_{1n}=k]$    &  $\P[S_{1n}\le k]$ \\
 &&&\\
  \hline \hline
 &&&\\
~0~&~0   &    33.490\% &  ~33.490\%   \\
~1~&10   &    40.188\% &  ~73.678\%   \\
~2~&20   &    20.094\% &  ~93.771\%   \\
~3~&30   &    ~5.358\% &  ~99.130\%   \\
~4~&40   &    ~0.804\% &  ~99.934\%   \\
~5~&50   &    ~0.064\% &  ~99.998\%   \\
~6~&60   &    ~0.002\% &  100.000\%   \\
&&&\\\hline
\end{tabular}
\end{center}
\end{table}

The expected total loss amount is given by $\E[L]=l\E[S_{1n}]=10$. We see that there is a 26.32\% probability (corresponding to $\P[S_{1n} > 10]=1-\P[S_{1n} \leq 10]$) that the company will turn out paying more than the expectation.
Thus, we cannot simply ask the expected loss as premium. This justifies the premium principle adopted in \S\;\ref{ss-techPremium}.

Now, we compute the cost of capital per policy given in~\eqref{eq: riskload} as a function of the number $N$ of policies in the portfolio. The results are displayed in Table~\ref{tbl-clt} for both risk measures VaR and TVaR, and when taking $p=1/6, \;1/4$ and $1/2$, respectively. The expected total loss amount $\E[L]=l\E[S_{1n}]=nlp$ will change accordingly.

\begin{table}[h]
\captionsetup{width=11cm} % To adjust the width of the caption to a given width
\begin{center}
\caption[Cost of capital] {\small \sf The Risk loading per policy as a function of the number $N$ of policies in the portfolio (with $n=6$).} \label{tbl-clt}
\begin{tabular}{c c@{\extracolsep{2em}} c@{\extracolsep{2em}} c@{\extracolsep{2em}} c}\hline
&&&&\\
&&  \multicolumn{3}{c}{\bf Risk Loading $R$ per policy} \\
{\bf Risk measure} &{\bf Number $N$}  & \multicolumn{3}{c}{\bf with probability}  \\
${\bf\rho}$  & {\bf of Policies}  &   $p=1/6$    & $p=1/4$ & $p=1/2$\\
  &&&&\\
 \hline\hline
 &   &  & & \\
{\bf VaR}&&&&\\
~& ~~~~~1   & 3.000 & 3.750 & 4.500 \\
~& ~~~~~5   & 1.500 & 1.650 & 1.800  \\
~& ~~~~10   & 1.050 & 1.200 & 1.350 \\
~& ~~~~50   & 0.450 & 0.540 & 0.600 \\
~& ~~~100   & 0.330 & 0.375 & 0.420 \\
~& ~1'000    & 0.102 & 0.117 & 0.135  \\
~& 10'000    & 0.032 & 0.037 & 0.043  \\
&&&&\\
{\bf TVaR}&&&&\\
~& ~~~~~1   & 3.226 & 3.945 & 4.500  \\
~& ~~~~~5   & 1.644 & 1.817 & 1.963 \\
~& ~~~~10   & 1.164 & 1.330 & 1.482 \\
~& ~~~~50   & 0.510 & 0.707 & 0.675 \\
~& ~~~100   & 0.372 & 0.425 & 0.476 \\
~& ~1'000    & 0.116 & 0.134 & 0.154  \\
~& 10'000    & 0.037 & 0.042 & 0.049  \\
&&&&\\\hline
&&&&\\
\multicolumn{2}{l} {$\E[L]/N$} & {\em 10.00} &{\em 15.00} &{\em 30.00}\\
&&&&\\\hline
\end{tabular}
\end{center}
\end{table}

Note that, when considering a large number $N$ of policies,  the binomial distribution of $L$ could be replaced by the normal distribution ${\bf \cal N}(Nnp, Nnp(1-p))$ (for $Nn\ge 30$ and $p$ not close to 0, nor 1; e.g. $np>5$ and $n(1-p)>5$) using the Central Limit Theorem (CLT). The VaR of order $\alpha$ of $L$ could then be deduced from the $\alpha$th-quantile, $q_{\alpha}$, of the standard normal distribution ${\bf \cal N}(0,1)$, as:
\begin{equation}\label{eq:GaussVaR}
VaR_{\alpha}(L)=\sqrt{Nnp(1-p)}~q_{\alpha}+Nnp
\end{equation}
Thus the risk loading $R$ would become, in the case of $\rho$ being VaR:
 \begin{equation*}
 R=\eta \times\sqrt{\frac{n l p(1-p)}{N}}~q_{\alpha}
 \end{equation*}
ever smaller as a function of $N$.

We can see in Table~\ref{tbl-clt} that the risk loading drops practically by a factor 100 for a portfolio of 10'000 policies, compared with the one computed for one policy ($N=1$) that represents 30\% of the loss expectation ($\E[L]=10$ in this case).
We notice also numerically that, if $R$ increases with $p$, the relative risk per policy $R/\E[L^{(1)}]$ decreases when $p$ increases. When considering the Gaussian approximation and the explicit VaR given in \eqref{eq:GaussVaR}, the relative risk per policy when choosing $\rho=$VaR, is, as a function of $p$, of the order $1/\sqrt{p}$, giving back the numerical result.
Finally, it is worth noting that the risk loading with TVaR is always slightly higher than with VaR for the same threshold, as TVaR goes beyond VaR in the tail of the distribution.

In this setting, a fair game is defined by having an equal probability of losing at each exposure: $p=1/n$. The biased game will be when the probability differs from $1/n$, generally bigger. We can thus define two states, one with a "normal" or equilibrium state ($p=1/n$) and a "crisis" state with a probability $q>\!>p$. In the next section, we will introduce this distinction.

 \subsection{Introducing a structure of dependence to reveal a systemic risk}

We propose two examples of models introducing a structure of dependence between the risks, in order to explore the occurrence of a systemic risk and, as a consequence, the limits to diversification.
We still consider  the sequence $(X_i, i=1,\ldots,Nn)$ to model the occurrence of the risk, with a given severity $l$, for $N$ policies, but do not assume anymore that the $X_i$'s are iid.

\subsubsection{A dependent model, but conditionally independent}\label{cdtional-indpdce}

We assume that the occurrence of the risks $X_i$'s depends on another phenomenon,  represented by a rv, say $U$. Depending on the intensity of the phenomenon, {\it i.e.} the values taken by $U$, a risk $X_i$ has more or less chances to occur. Suppose that the dependence between the risks is totally captured by $U$.
Consider, w.l.o.g., that $U$ can take two possible values denoted by 1 and 0; $U$ can then be modeled by a Bernoulli ${\cal B}(\tilde p)$, $0<\tilde p<\!< 1$.  The rv $U$ will be identified to the occurrence of a state of systemic risk. So if $\tilde p$ could mathematically  take any value between 0 and 1, but we choose it here to be very small since we want to explore rare events.
We still model the occurrence of the risks with a Bernoulli, but with a parameter depending on $U$. Since $U$ takes two possible values, the same holds for the parameters of the Bernoulli distribution of  the conditionally independent rv's $X_i\;|\;U$, namely
$$
X_i\;|\;(U=1) \sim {\cal B}(q) \quad \text{and}\quad X_i\;|\;(U=0) \sim {\cal B}(p)
$$
We choose $q>\!>p$, so that whenever $U$ occurs ({\it i.e.} $U=1$), it has a big impact in the sense that  there is a higher chance of loss. We include this effect in order to have a systemic risk (non-diversifiable) in our portfolio.

Looking at the total amount of losses $S_{Nn}$, its distribution can then be written, for $k\in \N$,  as
\begin{eqnarray*}
\P(S_{Nn}=k)&=&  \P[(S_{Nn}=k) ~ |~(U=1)] \P(U=1)+ \P[(S_{Nn}=k) ~ |~(U=0)] \P(U=0)\\
&=& \tilde p ~ \P[(S_{Nn}=k) ~ |~(U=1)] ~+~ (1-\tilde p) ~ \P[(S_{Nn}=k) ~ |~(U=0)]
\end{eqnarray*}
The conditional and independent variables, $\tilde S_q:=S_{Nn} |~(U=1)$ and $\tilde S_p:=S_{Nn} |~(U=0)$, are distributed as Binomials ${\cal B}(Nn,q)$ and ${\cal B}(Nn,p)$, with mass probability distributions denoted by $f_{\tilde S_q}$ and  $f_{\tilde S_p}$ respectively.
The mass probability distribution $f_S$ of $S_{Nn}$  appears as a mixture of $f_{\tilde S_q}$ and  $f_{\tilde S_p}$ (see e.g. \cite{Ev:Ha}): 
\begin{equation}\label{df-mixed1}
f_S  = ~ \tilde p ~f_{\tilde S_q}~+~ (1-\tilde p)~f_{\tilde S_p}  \quad\text{with}~\tilde S_q\sim {\cal B}(Nn,q) ~\text{and}~\tilde S_p\sim {\cal B}(Nn,p)
\end{equation}
Note that $\tilde p=0$ gives back the normal state, developed in \S\;\ref{iid}.

The expected loss amount for the portfolio, denoting $L=L^{(N)}$, is given by
\begin{equation*}
 \E[L] =l\times\E[S_{Nn}]= l\times \Big(\tilde p ~\E[\tilde S_q]~+~ (1-\tilde p)~\E[\tilde S_p]\Big)  = Nnl~\Big( \tilde p ~q~+~(1-\tilde p) ~p \Big)
\end{equation*}
whereas for each policy, it is
\begin{equation}\label{eq:expect2}
\frac{l}{N}\E[L]~=~ l~n~\Big( \tilde p ~q~+~(1-\tilde p) ~p \Big)
\end{equation}
from which we deduce the risk loading defined in  \eqref{eq:RAC} and \eqref{eq: riskload}.

Let us evaluate the variance $var(S_{Nn})$ of $S_{Nn}$. Straightforward computations (see \cite{3M}) give
$$
\E[S_{Nn}^2] = Nn \left[\tilde p ~q \big(1-q+ Nnq\big) ~+~ (1-\tilde p)~p \big(1-p+ Nnp\big)  \right]
$$
which, combined with~\eqref{eq:expect2}, provides
\begin{equation*}
var(S_{Nn}) = Nn \left[q(1-q)\tilde p + p(1-p) (1-\tilde p)+ Nn(q-p)^2 \tilde p(1-\tilde p)\right]
\end{equation*}
from which we deduce the variance for the loss of one contract as $ \frac1{N^2}~var(L)=\frac{l^2}{N^2}~var(S_{Nn}) $, {i.e.}
\begin{equation}\label{eq:var2}
\frac1{N^2}~var(L)= \frac{l^2 n}{N} \Big( q(1-q)\tilde p + p(1-p) (1-\tilde p) \Big) ~+~ l^2 n^2 (q-p)^2 \tilde p(1-\tilde p)
\end{equation}
Notice that in the variance for one contract, the first term will decrease as the number of contracts increases, but not the second one. It does not depend on $N$ and thus represents the non-diversifiable part of the risk.

\begin{table}[h]
\small
\begin{center}
\caption[Risk Loading for non-diversifiable risk]{\small \sf For Model \eqref{df-mixed1}, the Risk loading per policy as a function of the probability of occurrence of a systemic risk in the portfolio using VaR and TVaR measures with $\alpha=99\%$. The probability of giving a loss in a state of systemic risk is chosen to be $q=50\%$.}\label{tbl-div2}
\begin{tabular}{c c@{\extracolsep{2em}} c@{\extracolsep{1em}} c@{\extracolsep{1em}} c@{\extracolsep{1em}} c@{\extracolsep{1em}} c}\hline
&&&&&&\\
 {\bf Risk measure} & {\bf Number $N$}    &         \multicolumn{4}{c}{\bf Risk Loading $R$} & \\
    ${\bf\rho}$    &{\bf of Policies} & {\bf in a normal state} &     \multicolumn{4}{c}{{\bf with occurrence of a crisis state}}           \\
   & &   $\tilde{p}=0$    & $\tilde{p}=0.1\%$ & $\tilde{p}=1.0\%$ & $\tilde{p}= 5.0\%$ & $\tilde{p}=10.0\%$ \\
  &&&&&&\\
 \hline\hline
 &     & & \\
{\bf VaR}&&&&&&\\
~& ~~~~~1   & 3.000 & 2.997 & 4.469 & 4.346 & 5.693 \\
~& ~~~~~5   & 1.500 & 1.497 & 2.070 & 3.450 & 3.900 \\
~& ~~~~10   & 1.050 & 1.047 & 1.770 & 3.300 & 3.450 \\
~& ~~~~50   & 0.450 & 0.477 & 1.410 & 3.060 & 3.030 \\
~& ~~~100   & 0.330 & 0.327 & 1.605 & 3.000 & 2.940 \\
~& ~1'000   & 0.102 & 0.101 & 2.549 & 2.900 & 2.775 \\
~& 10'000   & 0.032 & 0.029 & 2.837 & 2.866 & 2.724 \\
&&&&&&\\
{\bf TVaR}&&&&&&\\
~& ~~~~~1   & 3.226 & 3.232 & 4.711 & 4.755 & 5.899 \\
~& ~~~~~5   & 1.644 & 1.707 & 2.956 & 3.823 & 4.146 \\
~& ~~~~10   & 1.164 & 1.266 & 2.973 & 3.578 & 3.665 \\
~& ~~~~50   & 0.510 & 0.760 & 2.970 & 3.196 & 3.141 \\
~& ~~~100   & 0.372 & 0.596 & 2.970 & 3.098 & 3.020 \\
~& ~1'000   & 0.116 & 0.396 & 2.970 & 2.931 & 2.802 \\
~& 10'000   & 0.037 & 0.323 & 2.970 & 2.876 & 2.732 \\
&&&&&&\\\hline
&&&&&&\\
\multicolumn{2}{l} {$\E[L]/N$} & {\em 10.00} &{\em 10.02} &{\em 10.20}&{\em 11.00}& {\em 12.00}\\
&&&&&&\\\hline
\end{tabular}
\end{center}
\end{table}

{\it Numerical application.}

For this application, we keep the same parameters $n=6$ and $p=1/n$ as in \S~\ref{iid} , and we choose the loss probability during the crisis to be $q=1/2$. We explore different probabilities  $\tilde p$ of occurrence of a crisis. The calculation consists in mixing the two Binomial distributions, according to \eqref{df-mixed1}, for an increasing number of policies $N$. Results for the two choices of risk measures are shown in Table~\ref{tbl-div2}.

In Table~\ref{tbl-div2}, we see well the effect of the non-diversifiable risk. As expected, when the probability of occurrence of a crisis is high, the diversification does not play a significant role anymore already with 100 contracts in the portfolio. The interesting point is that for $\tilde p \ge 1\%$, the risk loading barely changes when there is a large number of policies (starting at $N=1000$) in the portfolio. This is true for both VaR and TVaR. The non-diversifiable term dominates the risk. When looking at a lower probability $\tilde p$ of occurrence of a crisis, we notice that the choice of the risk measure matters. For instance, when choosing $\tilde{p}=0.1\%$,  the risk loading, compared to the normal state, is multiplied by 10 in the case of TVaR, for $N=10'000$ policies, and hardly moves in the case of VaR! This effect remains, but to a lower extend, when diminishing the number of policies. It is clear that the VaR measure does not capture well the crisis state, while TVaR is sensitive to the change of state, even with such a small probability and a high number of policies.

\subsubsection{A more realistic setting to introduce a systemic risk}

We adapt further the previous setting to a more realistic description of a crisis. At each of the $n$ exposures to the risk, in a state of systemic risk, the entire portfolio will be touched by the same increased probability of loss, whereas, in a normal state, the entire portfolio will be subject to the same equilibrium probability of loss.

For this modeling, it is more convenient to rewrite the sequence $(X_{i},\; i=1,\ldots,Nn)$ with a vectorial notation, namely $(\mathbf{X}_j,\; j=1,\ldots,n)$ where the vector
$\mathbf{X}_j$ is defined by $\mathbf{X}_j=(X_{1j},\ldots, X_{Nj})^T$.
Hence the total loss amount $S_{Nn}$ can be rewritten as
$$
S_{Nn}=  \sum_{j=1}^n \tilde S^{(j)} \quad\text{where}\quad \tilde S^{(j)}\;\text{is the sum of the components of}\; \mathbf{X}_j :\;\tilde S^{(j)}= \sum_{i=1}^N X_{ij}
$$
We keep the same notation for the Bernoulli rv $U$ determining the state and for its parameter $\tilde p$. But now, instead of defining a normal ($U=0$) or a crisis ($U=1$) state  on each element of $(X_{i},\; i=1,\ldots,Nn)$, we do it on each vector $\mathbf{X}_j$, $1\le j\le n$.
It comes back to define a sequence of iid rv's $(U_j, \, j=1,\ldots,n)$ with parent rv $U$.
Hence we deduce that $\tilde S^{(j)}$ follows a Binomial distribution whose probability depends on $U_j$:
$$
\tilde S^{(j)}\;|\;(U_j=1) \sim {\cal B}(N,q) \quad \text{and}\quad \tilde S^{(j)}\;|\;(U_j=0) \sim {\cal B}(N,p)
$$
Note that these conditional rv's are independent.

Let us introduce the event $A_l$ defined, for $ l=0,\ldots,n$, as
$$
A_l:=\{l \;\text{vectors} \; \mathbf{X}_j \;\text{are exposed to a crisis state and}\;  n\!-\!l \;\text{to a normal state}\}=\Big(\sum_{j=1}^n U_j=l \Big)
$$
whose probability is given by
$\displaystyle \P(A_l)=\P\Big(\sum_{j=1}^n U_j=l \Big) = \binom{n}{l}\; \tilde p^l \; (1-\tilde p)^{n-l} $.\\
We can then write that
\begin{equation} \label{loss_vec}
 \P(S_{Nn}=k)  = \sum_{l=0}^n \P(S_{Nn}=k\; |\; A_l)\P(A_l) =
 \sum_{l=0}^n ~\binom{n}{l} \tilde p^l ~ (1-\tilde p)^{n-l}~\P\big[\tilde S_q^{(l)}+\tilde S_p^{(n-l)}=k \big]
 \end{equation}
 with, by conditional independence,
 \begin{equation}\label{partial_sums}
\tilde S_q^{(l)}=\!\sum_{j=1}^{l}  \Big( \tilde S^{(j)}\;|\; U_j\!=\!1\Big)\sim {\cal B}(Nl,q) ~\;\text{and}~\; \tilde S_p^{(n-l)}=\! \sum_{j=1}^{n-l}  \Big( \tilde S^{(j)}\;|\; U_j\!=\!0\Big)\sim {\cal B}(N(n-l),p) \quad
\end{equation}
Expectation and variance are obtained by straightforward computations (see \cite{3M}). We have
\begin{eqnarray}\label{expect3}
\E[S_{Nn}]= Nn\left(\big(q-p\big)  \tilde p + p\right)
\end{eqnarray}
and, for one contract,
\begin{equation*}
\frac1{N}\E[L]=\frac{l}{N}\E[S_{Nn}] % = nl\left(\tilde p (q-p)+ p\right)
= nl\left(\tilde p ~q~+ (1-\tilde p) ~p\right)
\end{equation*}
which is equal to the expectation \eqref{eq:expect2} obtained with the previous method. The variance can be deduced from \eqref{expect3} and
\begin{eqnarray*}
\E[S_{Nn}^2]= N n\Big(q(1-q)\tilde p + p(1-p) (1-\tilde p)\Big)+ ~ N^2n^2\Big(p(1-\tilde p)+q\tilde p\Big)^2+ N^2n(q-p)^2 \tilde p(1-\tilde p)
\end{eqnarray*}
hence
$$
var(S_{Nn}) = Nn \left[q(1-q)\tilde p + p(1-p) (1-\tilde p)+ N(q-p)^2 \tilde p(1-\tilde p)\right]
$$
which is different from the variance $var(S_{Nn})$ obtained with the previous model in \S\;\ref{cdtional-indpdce}. \\
Now for one contract we obtain:
\begin{equation}\label{eq:var3}
\frac1{N^2}var(L)=\frac{l^2}{N^2}var(S_{Nn}) =  \frac{l^2n}{N}\Big(q(1-q)\tilde p + p(1-p) (1-\tilde p)\Big)~+~l^2n~ (q-p)^2 \tilde p(1-\tilde p)
\end{equation}
Notice that the last term appearing in \eqref{eq:var3} is only multiplied by $n$ and not $n^2$ as in \eqref{eq:var2}, and not diversifiable by the number $N$ of policies.  It looks alike the one of \eqref{eq:var2}, however its effect is smaller than in the previous model.
With this method we have also achieved to produce a process with a non-diversifiable risk.

\newpage

{\it Numerical application.}

Let us revisit our numerical example. In this case, we cannot, contrary to the previous cases, directly use an explicit expression for the distributions. We have to go through Monte-Carlo simulations.
At each of the $n$ exposures to the risk, we first have to choose between a normal or a crisis state. Since, we take here $n=6$, the chances of choosing a crisis state when $\tilde p= 0.1$\% is very small. To get enough of the crisis states, we need to do enough simulations, and then average over all the simulations. The results shown in Table~\ref{tbl-div3} are obtained with 10 million simulations. We ran it also with 1 and 20 million simulations to check the convergence. It converges well as can be seen in Table~\ref{tbl-c3-conv}.

\begin{table}[h]
\small
\begin{center}
\caption[Risk Loading for non-diversifiable risk]{\small \sf For Model \eqref{loss_vec}, the Risk loading per policy as a function of the probability of occurrence of a systemic risk in the portfolio using VaR and TVaR measures with $\alpha=99\%$. The probability of giving a loss in a state of systemic risk is chosen to be $q=50\%$.}\label{tbl-div3}
\begin{tabular}{c c@{\extracolsep{2em}} c@{\extracolsep{1em}} c@{\extracolsep{1em}} c@{\extracolsep{1em}} c@{\extracolsep{1em}} c}\hline
&&&&&&\\
 {\bf Risk measure} & {\bf Number $N$}    &         \multicolumn{4}{c}{\bf Risk Loading $R$} & \\
    ${\bf\rho}$    &{\bf of Policies} & {\bf in a normal state} &  \multicolumn{4}{c}{{\bf with occurrence of a crisis state}} \\
       & &   $\tilde{p}=0$    & $\tilde{p}=0.1\%$ & $\tilde{p}=1.0\%$ & $\tilde{p}= 5.0\%$ & $\tilde{p}=10.0\%$ \\
  &&&&&&\\
 \hline\hline
 &     & & \\
{\bf VaR}&&&&&&\\
~& ~~~~~~1   & 3.000 & 2.997 & 2.969 & 4.350 & 4.200 \\
~& ~~~~~~5   & 1.500 & 1.497 & 1.470 & 1.650 & 1.800 \\
~& ~~~~~10   & 1.050 & 1.047 & 1.170 & 1.350 & 1.500 \\
~& ~~~~~50   & 0.450 & 0.477 & 0.690 & 0.990 & 1.200 \\
~& ~~~~100   & 0.330 & 0.357 & 0.615 & 0.945 & 1.170 \\
~& ~~1'000   & 0.102 & 0.112 & 0.517 & 0.882 & 1.186 \\
~& ~10'000   & 0.032 & 0.033 & 0.485 & 0.860 & 1.196 \\
~& 100'000   & 0.010 & 0.008 & 0.475 & 0.853 & 1.199 \\
&&&&&&\\
{\bf TVaR}&&&&&&\\
~& ~~~~~~1   & 3.226 & 3.232 & 4.485 & 4.515 & 4.448 \\
~& ~~~~~~5   & 1.644 & 1.792 & 1.870 & 2.056 & 2.226 \\
~& ~~~~~10   & 1.164 & 1.252 & 1.342 & 1.604 & 1.804 \\
~& ~~~~~50   & 0.510 & 0.588 & 0.824 & 1.183 & 1.408 \\
~& ~~~~100   & 0.375 & 0.473 & 0.740 & 1.118 & 1.358 \\
~& ~~1'000   & 0.116 & 0.348 & 0.605 & 1.013 & 1.295 \\
~& ~10'000   & 0.037 & 0.313 & 0.563 & 0.981 & 1.276 \\
~& 100'000   & 0.012 & 0.301 & 0.550 & 0.970 & 1.269 \\
&&&&&&\\\hline
&&&&&&\\
\multicolumn{2}{l} {$\E[L]/N$} & {\em 10.00} &{\em 10.02} &{\em 10.20}&{\em 11.00}& {\em 12.00}\\
&&&&&&\\\hline
\end{tabular}
\end{center}
\end{table}

The results shown in Table~\ref{tbl-div3} follow what we expect. The diversification due to the total number of policies is more effective for this model than for the previous one, but we still experience a part which is not diversifiable. We have also computed the case with 100'000 policies since we used Monte Carlo simulations. It is interesting to note that, as expected, the risk loading in the normal state continues to decrease. In this state, it decreases by $\sqrt{10}$. However, except for $\tilde{p}=0.1\%$ in the VaR case, the decrease becomes very slow when we allow for a crisis state to occur. The behavior of this model is more complex than the previous one, but more realistic, and we reach also the non-diversifiable part of the risk. For a high probability of occurrence of a crisis (1 every 10 years), the limit with VaR is reached already at 100 policies, while, with TVaR, it continues to slowly decrease.

Concerning the choice of risk measure, we see a similar behavior as in Table~\ref{tbl-div2} for the case $N=10'000$ and $\tilde{p}=0.1\%$: VaR is unable to catch the possible occurrence of a crisis state, which shows its limitation as a risk measure.
Although we know that there is a part of the risk that is non-diversifiable, VaR does not catch it really when $N=10'000$ or $100'000$  while TVaR does not decrease significantly between $10'000$ and $100'000$ reflecting the fact that the risk cannot be completely diversified away.

Finally, to explore the convergence of the simulations, we present in Table~\ref{tbl-c3-conv} the results obtained for $N=100$ and for various number of simulations.

\begin{table}[h]
\small
\begin{center}
\caption[Convergence of the Risk Loading for Model \eqref{loss_vec}]{\small \sf Testing the numerical convergence: the Risk loading as a function of the number of Monte Carlo simulations, for $N=100$, Model \eqref{loss_vec}, and the same parameters as in Table \ref{tbl-div3}.}\label{tbl-c3-conv}
\begin{tabular}{c c@{\extracolsep{2em}} c@{\extracolsep{1em}} c@{\extracolsep{1em}} c@{\extracolsep{1em}} c@{\extracolsep{1em}} c}\hline
&&&&&&\\
 {\bf Risk measure} & {\bf Number $N$}    &         \multicolumn{4}{c}{\bf Risk Loading $R$} & \\
    ${\bf\rho}$    &{\bf of Policies} & {\bf in a normal state} &  \multicolumn{4}{c}{{\bf with occurrence of a crisis state}} \\
   & &   $\tilde{p}=0$    & $\tilde{p}=0.1\%$ & $\tilde{p}=1.0\%$ & $\tilde{p}= 5.0\%$ & $\tilde{p}=10.0\%$ \\
  &&&&&&\\
 \hline\hline
 &     & & \\
{\bf VaR}&&&&&&\\
~& ~1 million & 0.330 & 0.357 & 0.615 & 0.945 & 1.170 \\
~& 10 million & 0.330 & 0.357 & 0.615 & 0.945 & 1.155 \\
~& 20 million & 0.330 & 0.357 & 0.615 & 0.945 & 1.170 \\
&&&&&&\\
{\bf TVaR}&&&&&&\\
~& ~1 million & 0.375 & 0.476 & 0.738 & 1.115 & 1.358 \\
~& 10 million & 0.374 & 0.472 & 0.739 & 1.117 & 1.357 \\
~& 20 million & 0.375 & 0.473 & 0.740 & 1.118 & 1.358 \\
&&&&&&\\\hline
&&&&&&\\
\multicolumn{2}{l} {$\E[L]/N$} & {\em 10.00} &{\em 10.02} &{\em 10.20}&{\em 11.00}& {\em 12.00}\\
&&&&&&\\\hline
\end{tabular}
\end{center}
\end{table}

It appears that, already for this number of policies ($N=100$), the number of simulations has no influence. Obviously, with a lower number of policies, the number of simulations plays a more important role as one would expect, while for a higher number of policies, it is insensitive to the number of simulations above 1 million.

\section{Comparison and discussion}\label{sec-discuss}

Let us start with the following table, presenting a summary of the expectation and the variance of the total loss amount per policy, obtained for each model.

\begin{table}[h]
\begin{center}
\caption[Summary of the analytical results]{\sf Summary of the analytical results (expectation and variance per policy) for the 3 models with cdf defined by \eqref{eq:cdf-L}, \eqref{df-mixed1} and  \eqref{loss_vec}, respectively.}\label{tbl-cases}
\begin{tabular}{c@{\extracolsep{2em}} c@{\extracolsep{2em}}c}\hline
&&\\
Model & Expectation $\displaystyle \frac1N \E[L]$ & Variance $\displaystyle \frac1{N^2}var(L)$\\
 &&\\\hline\hline
 &&\\
\eqref{eq:cdf-L} & $ln~p$ & $\displaystyle \frac{l^2n}{N}~p(1-p)$ \qquad \qquad \qquad \qquad \\[2ex]
\eqref{df-mixed1} & $ln~\Big( \tilde p ~q~+~(1-\tilde p) ~p \Big)$ & $\displaystyle \frac{l^2 n}{N} \Big( q(1-q)\tilde p + p(1-p) (1-\tilde p) \Big) +\mathbf{ l^2 n^2(q-p)^2 \tilde p(1-\tilde p)}$ \\[2ex]
\eqref{loss_vec} & $ln~\Big( \tilde p ~q~+~(1-\tilde p) ~p \Big)$ & $\displaystyle \frac{l^2 n}{N} \Big( q(1-q)\tilde p + p(1-p) (1-\tilde p) \Big) +\mathbf{ l^2 n~(q-p)^2 \tilde p(1-\tilde p)}$ \\
&&\\\hline
\end{tabular}
\end{center}
\end{table}

For the first model \eqref{eq:cdf-L}, we see that the variance decreases with increasing $N$, while for both other models \eqref{df-mixed1}  and \eqref{loss_vec} , the variance contains a term that does not depend on $N$, which corresponds to the presence of a systemic risk, and is not diversifiable.
Note that the variance for Model \eqref{df-mixed1} contains a non-diversifiable part that corresponds to $n$ times the non-diversifiable part of the variance for Model \eqref{loss_vec}.
This is consistent with the numerical results in Tables~\ref{tbl-div2}~and~\ref{tbl-div3}; indeed the smaller the non-diversifiable part, the longer the decrease of the risk loading $R$ ({\it  i.e.} effect of diversification) with increase of number of policies.
The latter model is the most interesting because it shows both the effect of diversification and the effect of the non-diversifiable term in a more realistic way. It assumes the occurrence of states that are dangerous to the whole portfolio, which is characteristic of a state of crisis in the financial markets.
Thus it is more suitable to explore other properties and the limits of diversification in times of crisis.

Concerning the choice of risk measure, we have already noticed that there was an issue when evaluating the VaR with small $\tilde p$. There are other less obvious stability problems, as for instance the VaR in Model with cdf \eqref{df-mixed1} for  $\tilde p=1\%$. It starts to decrease with $N$ increasing, then raises again for large $N$, while the TVaR decreases with $N$ increasing, then stabilizes to a value whenever $N\ge 50$.
For $\tilde p=0.1\%$, in both models with cdf \eqref{df-mixed1} and \eqref{loss_vec} respectively, VaR is very close to the case \eqref{eq:cdf-L} without systemic risk, while TVaR starts to be significantly impacted already with 50 policies, indicating that the systemic risk appears mostly beyond the $99\%$ threshold.
Even if there is a part of the risk that is non-diversifiable, VaR, under certain circumstances, might not catch it (see  \cite{emmer&tasche}, Proposition~3.3).

\section{Conclusion}

In this study, we have shown the effect of diversification on the pricing of insurance risk through a first simple modeling. Then, for understanding and analyzing possible limitations to diversification benefits, we propose two alternative stochastic models, introducing dependence between risks by assuming the existence of an underlying systemic risk.
These models, defined with mixing distributions, allow for a straightforward analytical evaluation of  the impact of the non-diversifiable part, which appears in the close form expression of the variance.
We have adopted here purposely a probabilistic approach for modelling the dependence and the existence of systemic risk. It could be easily generalized to a time series interpretation by assigning a time-step to each exposure $n$. In the last model, the occurrence of the rv $U=1$ could then be identified to the time of crisis.

In real life, insurers have to pay special attention to the effects that can weaken the diversification benefits. For instance, in the case of motor insurance, the appearance of a hail storm will introduce a "bias" in the usual risk of accident due to a cause independent of the car drivers, which will hit a big number of cars at the same time and thus cannot be  diversified among the various policies. There are other examples in life insurance for instance with pandemic or mortality trend that would affect the entire portfolio and cannot be diversified away. Special care must be given to those risks as they will affect greatly the risk loading of the premium as can be seen in our examples. These examples might also find applications for real cases. This approach can be generalized to investments and banking; both are subject to systemic risk, although of different nature than in the above insurance examples.

The last model we suggested, introducing the occurrence of crisis, may find an interesting application for investment and the change in risk appetite of investors. It will be the subject of a following paper.
Moreover, the models we introduce here allow to point out explicitly  to the impact of dependence; they are simple enough to compute analytic expression and analyze the impact of the emergence of systemic risks. 
Yet, they are formulated in such a way that extensions to more sophisticated models, are easy and clear. In particular, it makes possible to obtain an extension to non identically distributed rv's,  or  when considering random severity. Another interesting perspective would be to consider econometric models with multiple states.

{\bf Acknowledgment:} Partial support from the project RARE - 318984 (a Marie Curie IRSES Fellowship within the 7th European Community Framework Program) is kindly acknowledged.

\end{document}